\documentstyle[12pt]{article}

\catcode`\@=11

\def\d{\hbox{\rm d}}
\def\diag{\mathop{\hbox{\,\rm diag}}\nolimits}
\def\tr{\mathop{\hbox{\,\rm tr}}\nolimits}
\def\re{\mathop{\hbox{\,\rm Re}}\nolimits}
\def\im{\mathop{\hbox{\,\rm Im}}\nolimits}
\def\hr{\hbox{\bf R}}
\def\@{@}
\def\I{\hbox{\rm i}}
\def\D{\displaystyle}

\def\zero{0}
\def\one{1}
\def\innerp#1#2#3{\if #3\zero\langle\Phi_{#1},\Phi_{#2}\rangle
 \else\if #3\one \langle\Phi_{#1},\Lambda\Phi_{#2}\rangle\else
 \langle\Phi_{#1},\Lambda^{#3}\Phi_{#2}\rangle\fi\fi}

\def\aligned#1{\begin{array}{#1}}
\def\endaligned{\end{array}}

\def\matrix#1{\begin{array}{#1}}
\def\endmatrix{\end{array}}
\def\pmatrix#1{\left(\begin{array}{#1}}
\def\endpmatrix{\end{array}\right)}

\def\setboxz@h{\setbox\z@\hbox}
\def\wdz@{\wd\z@}
\def\boxz@{\box\z@}
\newif\ifmsbmloaded@
\def\widetilde#1{
\ifmsbmloaded@
  \setboxz@h{$\m@th#1$}\ifdim\wdz@>\tw@ em\mathaccent"0\msbfam@5D{#1}\else
  \mathaccent"0365{#1}\fi\else\mathaccent"0365{#1}\fi}

\newtheorem{theorem}{\bf Theorem}
\newtheorem{lemma}{\bf Lemma}

\newtheorem{remark}{\sl Remark}

\def\wd@ne{\wd\@ne}
\def\setbox@ne{\setbox\@ne}
\def\binrel@#1{\setboxz@h{\thinmuskip0mu
  \medmuskip\m@ne mu\thickmuskip\@ne mu$#1\m@th$}%
 \setbox@ne\hbox{\thinmuskip0mu\medmuskip\m@ne mu\thickmuskip
  \@ne mu${}#1{}\m@th$}%
 \setbox\tw@\hbox{\hskip\wd@ne\hskip-\wdz@}}
\def\overset#1\to#2{\binrel@{#2}\ifdim\wd\tw@<\z@
 \mathbin{\mathop{\kern\z@#2}\limits^{#1}}\else\ifdim\wd\tw@>\z@
 \mathrel{\mathop{\kern\z@#2}\limits^{#1}}\else
 {\mathop{\kern\z@#2}\limits^{#1}}{}\fi\fi}
\def\underset#1\to#2{\binrel@{#2}\ifdim\wd\tw@<\z@
 \mathbin{\mathop{\kern\z@#2}\limits_{#1}}\else\ifdim\wd\tw@>\z@
 \mathrel{\mathop{\kern\z@#2}\limits_{#1}}\else
 {\mathop{\kern\z@#2}\limits_{#1}}{}\fi\fi}

\def\voidtoken{}

\def\refby{}
\def\refpaper{}
\def\refjour{}
\def\refvol{}
\def\refpage{}
\def\refpages{}
\def\refyr{}
\def\refbook{}
\def\refpubl{}
\def\refpubladdr{}

\def\bibref{\footnotesize
  \baselineskip=8pt
  \global\def\refby{}
  \global\def\refpaper{}
  \global\def\refjour{}
  \global\def\refvol{}
  \global\def\refpage{}
  \global\def\refpages{}
  \global\def\refyr{}
  \global\def\refbook{}
  \global\def\refpubl{}
  \global\def\refpubladdr{}
  }
\def\by#1{\global\def\refby{#1}}
\def\paper#1{\global\def\refpaper{#1}}
\def\jour#1{\global\def\refjour{#1}}
\def\vol#1{\global\def\refvol{#1}}
\def\page#1{\global\def\refpage{#1}}

\def\yr#1{\global\def\refyr{#1}}
\def\book#1{\global\def\refbook{#1}}
\def\publ#1{\global\def\refpubl{#1}}

\def\endbibref{{
\ifx\refby\voidtoken \else \refby\fi
\ifx\refjour\voidtoken \else , \refjour\fi
\ifx\refbook\voidtoken \else , \refbook\fi
\ifx\refpubl\voidtoken \else , \refpubl\fi
\ifx\refvol\voidtoken \else \ \refvol\fi
\ifx\refpage\voidtoken \else , \refpage\fi
\ifx\refyr\voidtoken \else \ (\refyr)\fi.\vskip2pt}}

\catcode`\@=\active

\title{Finite dimensional integrable Hamiltonian
systems associated with DSI equation by Bargmann constraints}

\author{\small Zixiang Zhou\\
\small Institute of Mathematics, Fudan University, Shanghai
200433, China\\
\small E-mail: zxzhou\@guomai.sh.cn\\
\small Wen-Xiu Ma\\
\small Department of Mathematics, City University of Hong Kong,
Kowloon, Hong Kong, China\\
\small E-mail: mawx\@cityu.edu.hk}
\date{}

\begin{document}
\maketitle

\begin{abstract}
The Davey-Stewartson I equation is a typical integrable 
equation in 2+1 dimensions. Its Lax system being essentially 
in 1+1 dimensional form has been found through nonlinearization 
from 2+1 dimensions to 1+1 dimensions. In the present paper,
this essentially 1+1 dimensional Lax system is further
nonlinearized into 1+0 dimensional Hamiltonian systems by taking
the Bargmann constraints. It is shown that the resulting 1+0
dimensional Hamiltonian systems are completely integrable in
Liouville sense by finding a full set of integrals of motion and
proving their functional independence.
\end{abstract}

\section{Introduction}

The Davey-Stewartson I (DSI) equation is a famous 2+1 dimensional
integrable equation which describes the motion of water wave
\cite{bib:DS}. This equation has localized soliton solutions and
has been studied in various ways, such as inverse scattering
\cite{bib:Boiti, bib:Fokas}, binary Darboux transformation
\cite{bib:MS}, nonlinearization to 1+1 dimensional problems
\cite{bib:Zhou2n, bib:Zhou2narb} etc.

For 1+1 dimensional integrable systems, the nonlinearization
procedure, both mono-non\-lineari\-zation \cite{bib:Cao1} and
binary nonlinearization \cite{MaS-PLA1994}, reduces them to finite
dimensional (1+0 dimensional) integrable Hamiltonian systems
\cite{bib:MaD, bib:Geng, bib:Zeng, bib:ZRG}. Therefore, it
transforms a partial differential equation to a systems of
ordinary differential equations. This greatly simplifies the
procedure of getting solutions, at least numerical solutions. Some
important explicit solutions, especially periodic or
quasi-periodic solutions have been obtained in this way.

The nonlinear constraint method has also been applied to some 2+1
dimensional equations like the KP, MKP, N-wave equations etc.
\cite{bib:Cao2, bib:Cao3, bib:MZNwave}. For the DSI equation, we have
already found its new Lax system (\ref{eq:lp}) by
nonlinearization in which all the derivatives are separated.
Each pair of equations in this system is 1+1 dimensional.
Hence the derived Lax system is looked as essentially 1+1
dimensional because we can use 1+1 dimensional method to solve
it. It is possible to nonlinearize this essentially 1+1
dimensional system again to get finite dimensional Hamiltonian
systems.

In the present paper, we show that there are Bargmann constraints
which reduce the DSI equation to finite dimensional Hamiltonian
systems. These Hamiltonian systems have a full set of 
integrals of motions and these integrals of motion are functionally
independent in a dense open subset of the phase space. Therefore,
these Hamiltonian systems are completely integrable in Liouville
sense.

\section{Nonlinearization}

We consider the following Lax system \cite{bib:Zhou2n}
\begin{equation}
   \aligned{l}
   \Phi_x=W^x\Phi=\left(\begin{array}{ccc}
   \I\lambda &0 &\I f\\
   0 &\I\lambda &\I g\\
   \I\bar f &\I\bar g &0
   \end{array}\right)\Phi\\
   \Phi_y=W^y\Phi=
   \left(\begin{array}{ccc}
   \I\lambda &u &\I f\\
   -\bar u &-\I\lambda &-\I g\\
   \I\bar f &-\I\bar g &0
   \end{array}\right)\Phi \\
   \Phi_t=W^t\Phi\\
   =\!\left(\!\!\!\!\begin{array}{l}
   -2\I\lambda^2\!+\!\I|u|^2\!+\!\I v_1\\2\bar u\lambda\!+\!\I\bar u_y\\
   -2\I\bar f\lambda\!+\!2\bar f_y
   \end{array}\!\!\!\!\!\!\!\!\!\!
   \begin{array}{c}
   -2u\lambda\!+\!\I u_y\\2\I\lambda^2\!-\!\I|u|^2\!-\!\I v_2\\
   2\I\bar g\lambda\!+\!2\bar g_y
   \end{array}\!\!\!\!\!\!\!\!\!\!
   \begin{array}{r}
   -2\I f\lambda\!-\!2f_y\\2\I g\lambda\!-\!2g_y\\-2\I(|f|^2\!-\!|g|^2)
   \end{array}\!\!\!\!\right)\!\Phi.
   \endaligned \label{eq:lp}
\end{equation}
Here $u$, $f$ and $g$ are complex functions, $v_1$ and $v_2$ are
real functions.

Its integrability conditions $\Phi_{xy}=\Phi_{yx}$,
$\Phi_{xt}=\Phi_{tx}$ and $\Phi_{yt}=\Phi_{ty}$
consist of the following three parts.

(1) DSI equation
\begin{equation}
   \aligned{l}
   -\I u_t=u_{xx}+u_{yy}+2|u|^2u+2(v_1+v_2)u\\
   v_{1,x}-v_{1,y}=v_{2,x}+v_{2,y}=-(|u|^2)_x.
   \endaligned \label{eq:DS}
\end{equation}

(2) Standard Lax pair of the DSI equation
\begin{equation}
   \aligned{l}
   F_y=\left(\begin{array}{ccc} 1 &0\\ 0 &-1\end{array}\right) F_x
    +\left(\begin{array}{ccc} 0 &u\\ -\bar u &0\end{array}\right) F\\
   F_t=2\I \left(\begin{array}{ccc} 1 &0\\ 0 &-1\end{array}\right) F_{xx}
    +2\I \left(\begin{array}{ccc} 0 &u\\ -\bar u &0\end{array}\right) F_x\\
    \qquad+\I\left(\begin{array}{ccc} |u|^2+2v_1 &u_x+u_y\\
    -\bar u_x+\bar u_y &-|u|^2-2v_2\end{array}\right) F
   \endaligned \label{eq:orglpDS}
\end{equation}
where $F=(f,g)^T$.

(3) Nonlinear constraint
\begin{equation}
   FF^*=\frac 12 \left(\begin{array}{ccc}
    v_1 &u_x\\ \bar u_x &v_2 \end{array}\right).
   \label{eq:1stconstrDS}
\end{equation}

Hence any solution of (\ref{eq:DS})--(\ref{eq:1stconstrDS}) gives
a solution of the DSI equation.

Notice that if $\Phi$ is a vector solution of (\ref{eq:lp}) for
real $\lambda$, then
$\Psi=\I\bar\Phi$ is a solution of the adjoint equations
\begin{equation}
   \aligned{l}
   \Psi_x=-(W^x)^T\Psi\qquad
   \Psi_y=-(W^y)^T\Psi\\
   \Psi_t=-(W^t)^T\Psi
   \endaligned
\end{equation}
where each entry of $\bar\Phi$ is the complex conjugation of the
corresponding entry of $\Phi$.

In order to obtain the finite dimensional Hamiltonian systems, we
first nonlinearize the $y$-equation of (\ref{eq:lp}) in the
following way.

Consider the pair $\Phi_y=W^y\Phi$ and $\Phi_t=W^t\Phi$. Let
$w=(u,-\bar u,\I f,\I\bar f,-\I g,-\I\bar g)$ containing all the
variables in $W^y$ \cite{bib:Geng}. Then the recursion relations
of this AKNS system can be expressed in Lenard form
\begin{equation}
   JG_l=KG_{l-1}\qquad (l=1,2,\cdots)
\end{equation}
where $(J,K)$ is the Lenard pair ($J$ is a non-de\-ge\-ne\-rate
constant matrix) which was given by \cite{bib:Geng} (here
$\alpha_1,\alpha_2,\alpha_3,\beta_1,\beta_2,\beta_3$ in \cite{bib:Geng} are
$\alpha_1=\I$, $\alpha_2=-\I$, $\alpha_3=0$, $\beta_1=-2\I$,
$\beta_2=2\I$, $\beta_3=0$) and $\{G_0,G_1,G_2,\cdots\}$ is the
Lenard sequence. The first element of this Lenard sequence is
given by \cite{bib:Geng} as 
\begin{equation}
   G_0=-2(-\bar u,u,\I\bar f,\I f,-\I\bar g,-\I g)
\end{equation}
which is a kernel of $K$.

On the other hand, the variation of spectral parameter can be
computed by the general formula \cite{bib:Ma}
\begin{equation}
   \aligned{rl}
   \D\frac{\delta\lambda}{\delta w}
    &\D=C_0\tr\left(\Phi\Psi^T\frac{\partial W^y}{\partial w}\right)\\
    &=C_0(\I\bar\phi_1\phi_2,\I\bar\phi_2\phi_1,
    \I\bar\phi_1\phi_3,\I\bar\phi_3\phi_1,
    \I\bar\phi_2\phi_3,\I\bar\phi_3\phi_2)
   \endaligned
\end{equation}
where $C_0$ is a constant.

Now let $\lambda_1,\cdots,\lambda_N$ be $N$ distinct non-zero real
numbers, $(\phi_{1\alpha},\phi_{2\alpha},\phi_{3\alpha})^T$ be the
corresponding solution of the Lax system (\ref{eq:lp}) for
$\lambda=\lambda_\alpha$. Let
$\Lambda=\diag(\lambda_1,\cdots,\lambda_N)$,
$\Phi_j=(\phi_{j1},\cdots,\phi_{jN})^T$.

By the general formulation of nonlinearization, we impose the
nonlinear constraint
\begin{equation}
   G_0=-2\sum_{j=1}^N\frac{\delta\lambda_j}{\delta w}
\end{equation}
which gives the relations
\begin{equation}
   \innerp 210=-\I u\qquad
   \innerp 310=f\qquad
   \innerp 320=-g
   \label{eq:2ndconstrDS}
\end{equation}
where $\langle V_1,V_2\rangle=V_1^*V_2$ for any two vectors
$V_1$ and $V_2$. These are the Bargmann constraints between
$(u,f,g)$ and $(\Phi_1,\Phi_2,\Phi_3)$.

\begin{remark}
If we consider the pair $\Phi_y=W^y\Phi$ and $\Phi_x=W^x\Phi$,
$G_0$ is different. In that case, we can not obtain Bargmann
constraints, but only Neumann constraints \cite{bib:ZMZ}.
\end{remark}

Let
\begin{equation}
   \aligned{l}
   L(\lambda)=\pmatrix{ccc} 1&&\\ &-1&\\ &&0\endpmatrix\\
   \qquad\D+\sum_{\alpha=1}^N\frac 1{\lambda-\lambda_\alpha}\pmatrix{ccc}
   \bar\phi_{1\alpha}\phi_{1\alpha}
    &\bar\phi_{2\alpha}\phi_{1\alpha}
    &\bar\phi_{3\alpha}\phi_{1\alpha}\\
   \bar\phi_{1\alpha}\phi_{2\alpha}
    &\bar\phi_{2\alpha}\phi_{2\alpha}
    &\bar\phi_{3\alpha}\phi_{2\alpha}\\
   \bar\phi_{1\alpha}\phi_{3\alpha}
    &\bar\phi_{2\alpha}\phi_{3\alpha}
    &\bar\phi_{3\alpha}\phi_{3\alpha}\endpmatrix.
   \endaligned
\end{equation}

\begin{lemma}\label{lemma:constr}
The Lax equations
\begin{equation}
   L_x=[W^x,L]\qquad L_y=[W^y,L]\qquad L_t=[W^t,L]
   \label{eq:Laxeq}
\end{equation}
hold if and only if (\ref{eq:2ndconstrDS}) holds.
\end{lemma}

\begin{demo}
Let $F_\alpha=(\phi_{1\alpha},\phi_{2\alpha},\phi_{3\alpha})^T$, then
\begin{equation}
   L(\lambda)=C+\sum_{\alpha=1}^N
   \frac 1{\lambda-\lambda_\alpha}F_\alpha F_\alpha^*.
\end{equation}
Since $F_{\alpha,y}=W^y(\lambda_\alpha)F_\alpha$ and
$(W^y(\lambda_\alpha))^*=-W^y(\lambda_\alpha)$, we have
\begin{equation}
   \aligned{rl}
   L_y(\lambda)&\D=\sum_\alpha\frac 1{\lambda-\lambda_\alpha}
    [W^y(\lambda_\alpha),F_\alpha F_\alpha^*]\\
   &\D=[W^y(\lambda),L(\lambda)-C]\\
   &\D\quad-\sum_\alpha\frac 1{\lambda-\lambda_\alpha}
    [W^y(\lambda)-W^y(\lambda_\alpha),F_\alpha F_\alpha^*]\\
   &\D=[W^y(\lambda),L(\lambda)-C]-\I\left[
    C,\sum_\alpha F_\alpha F_\alpha^*\right].
   \endaligned
\end{equation}
Hence $L_y(\lambda)=[W^y(\lambda),L(\lambda)]$ if and only if
\begin{equation}
   \I\left[C,\sum_\alpha F_\alpha F_\alpha^*\right]
   =[C,W^y(\lambda)].
\end{equation}
Written in the components, it becomes (\ref{eq:2ndconstrDS}).
If (\ref{eq:2ndconstrDS}) holds, the other two equations in
(\ref{eq:Laxeq}) are obtained similarly. The lemma is proved.
\end{demo}

From the above constraints, we have
\begin{equation}
   \aligned{l}
   f_x=\I\innerp 311
    +\I f(\innerp 330-\innerp 110)-ug\\
   g_x=-\I\innerp 321
    -\I g(\innerp 330-\innerp 220)-\bar uf\\
   f_y=\I\innerp 311
    +\I f(\innerp 330-\innerp 110)\\
   g_y=\I\innerp 321
    +\I g(\innerp 330-\innerp 220)\\
   u_x=2f\bar g\\
   u_y=-2\innerp 211+\innerp 210
   (\innerp 110-\innerp 220)\\
   v_1=2\innerp 130\innerp 310\qquad
   v_2=2\innerp 230\innerp 320
   \endaligned\label{eq:fxv2}
\end{equation}

With the constraints (\ref{eq:2ndconstrDS}), the system
(\ref{eq:lp}) is changed to a system of ordinary differential
equations
\begin{equation}
   \aligned{l}
   \Phi_{1,x}=\I\Lambda\Phi_1+\I f\Phi_3 \qquad
   \Phi_{2,x}=\I\Lambda\Phi_2+\I g\Phi_3 \\
   \Phi_{3,x}=\I\bar f\Phi_1+\I\bar g\Phi_2\\
   \Phi_{1,y}=\I\Lambda\Phi_1+u\Phi_2+\I f\Phi_3\\
   \Phi_{2,y}=-\bar u\Phi_1-\I\Lambda\Phi_2-\I g\Phi_3\\
   \Phi_{3,y}=\I\bar f\Phi_1-\I\bar g\Phi_2\\
   \Phi_{1,t}=(-2\I\Lambda^2+\I|u|^2+\I v_1)\Phi_1
   +(-2u\Lambda+\I u_y)\Phi_2\\
   \qquad+(-2\I f\Lambda-2f_y)\Phi_3\\
   \Phi_{2,t}=(2\bar u\Lambda+\I\bar u_y)\Phi_1
   +(2\I\Lambda^2-\I|u|^2-\I v_2)\Phi_2\\
   \qquad +(2\I g\Lambda-2g_y)\Phi_3\\
   \Phi_{3,t}=(-2\I\bar f\Lambda+2\bar f_y)\Phi_1
   +(2\I\bar g\Lambda+2\bar g_y)\Phi_2\\
   \qquad -2\I(|f|^2-|g|^2)\Phi_3
   \endaligned\label{eq:newlpDS}
\end{equation}
where $u$, $u_y$, $v_1$, $v_2$, $f$, $g$, $f_y$, $g_y$ are given
by (\ref{eq:2ndconstrDS}) and (\ref{eq:fxv2}).

$\re\phi_{j\alpha}$ and $\im\phi_{j\alpha}$ $(j=1,2,3;\,
\alpha=1,\cdots,N)$ form a system of coordinates of the phase
space $\hr^{6N}$. For simplicity, we use the complex coordinates
$\phi_{j\alpha}$ and $\bar\phi_{j\alpha}$ instead of
$\re\phi_{j\alpha}$ and $\im\phi_{j\alpha}$. $\hr^{6N}$ has the
standard symplectic form
\begin{eqnarray}
   \omega\D&=&2\sum_{j=1}^3\sum_{\alpha=1}^N
   \d\im(\phi_{j\alpha})\wedge \d\re(\bar\phi_{j\alpha})\nonumber\\
   &=&\I\sum_{j=1}^3\sum_{\alpha=1}^N
   \d\bar\phi_{j\alpha}\wedge \d\phi_{j\alpha}.
\end{eqnarray}
The corresponding Poisson bracket of two functions $\xi$ and $\eta$
is
\begin{equation}
   \{\xi,\eta\}=-\I\sum_{j=1}^{3}\sum_{\alpha=1}^N
   \left(\frac{\partial\xi}{\partial\phi_{j\alpha}}
   \frac{\partial\eta}{\partial\bar\phi_{j\alpha}}
   -\frac{\partial\xi}{\partial\bar\phi_{j\alpha}}
   \frac{\partial\eta}{\partial\phi_{j\alpha}}\right).
\end{equation}
By direct computation, we have

\begin{lemma}\label{lemma:H}
(\ref{eq:newlpDS}) is equivalent to three Hamiltonian equations
\begin{equation}
   \aligned{ll}
   \D\I\phi_{j\alpha,x}=\frac{\partial H^x}{\partial\bar\phi_{j\alpha}}\qquad
   &\D-\I\bar\phi_{j\alpha,x}=\frac{\partial H^x}{\partial\phi_{j\alpha}}\\
   \D\I\phi_{j\alpha,y}=\frac{\partial H^y}{\partial\bar\phi_{j\alpha}}\qquad
   &\D-\I\bar\phi_{j\alpha,y}=\frac{\partial H^y}{\partial\phi_{j\alpha}}\\
   \D\I\phi_{j\alpha,t}=\frac{\partial H^t}{\partial\bar\phi_{j\alpha}}\qquad
   &\D-\I\bar\phi_{j\alpha,t}=\frac{\partial H^t}{\partial\phi_{j\alpha}}
   \endaligned
\end{equation}
where
\begin{equation}
   \aligned{rl}
    H^x=&-\innerp 111
   -\innerp 221
   -|\innerp 130|^2\\
   &+|\innerp 230|^2\\
    H^y=&-\innerp 111
   +\innerp 221
   -|\innerp 130|^2\\
   &-|\innerp 230|^2
   -|\innerp 120|^2\\
    H^t=&2\innerp 112
    -2\innerp 222\nonumber\\
   &+4\re(\innerp 130
    \innerp 311)\nonumber\\
    &+4\re(\innerp 23\rangle
    \innerp 321)\nonumber\\
    &+4\re(\innerp 120
    \innerp 211) \\
    &+2(\innerp 330-\innerp 110)
    |\innerp 130|^2\nonumber\\
    &-2(\innerp 330-\innerp 220)
    |\innerp 230|^2\nonumber\\
    &-(\innerp 110-\innerp 220)
    |\innerp 120|^2.
   \endaligned \label{eq:HBarg}
\end{equation}
\end{lemma}

\section{Integrability}

\begin{lemma}\label{lemma:invol}
For any two complex numbers $\lambda$, $\mu$ and any positive
integers $k$, $l$,
\begin{equation}
   \{\tr L^k(\lambda),\tr L^l(\mu)\}=0.
\end{equation}
\end{lemma}

\begin{demo}
Denote $e_j=(0,\cdots,0,\underset{\scriptstyle
j}\to{1},0,\cdots,0)^T$, then
\begin{equation}
   \frac{\partial L(\lambda)}{\partial\bar\phi_{j\alpha}}
    =\frac 1{\lambda-\lambda_\alpha}F_\alpha e_j^T\qquad
   \frac{\partial L(\lambda)}{\partial\phi_{j\alpha}}
    =\frac 1{\lambda-\lambda_\alpha}e_j F_\alpha^*.
\end{equation}
Since
\begin{equation}
   \aligned{l}
   \D\tr\left(\frac{\partial L^k(\lambda)}{\partial\phi_{j\alpha}}\right)
    =\tr\sum_{r=1}^k
   L^{r-1}(\lambda)\frac{\partial L(\lambda)}{\partial\phi_{j\alpha}}
    L^{k-r}(\lambda)\\
   \D=k\tr\left(L^{k-1}(\lambda)
    \frac{\partial L(\lambda)}{\partial\phi_{j\alpha}}\right),\\
   \D\tr\left(\frac{\partial L^k(\lambda)}{\partial\bar\phi_{j\alpha}}\right)
   =k\tr\left(L^{k-1}(\lambda)
    \frac{\partial L(\lambda)}{\partial\bar\phi_{j\alpha}}\right),
   \endaligned
\end{equation}
we have
\begin{equation}
   \aligned{l}
   \D\quad \frac{\I}{kl}\{\tr L^k(\lambda),\tr L^l(\mu)\}\\
   \D=\sum_{j,\alpha}\bigg(\tr(L^{k-1}(\lambda)
    \frac{\partial L(\lambda)}{\partial\phi_{j\alpha}})
    \tr(L^{l-1}(\mu)\frac{\partial L(\mu)}{\partial\bar\phi_{j\alpha}})\\
   \D\qquad-\tr(L^{l-1}(\mu)\frac{\partial L(\mu)}{\partial\phi_{j\alpha}})
    \tr(L^{k-1}(\lambda)
    \frac{\partial L(\lambda)}{\partial\bar\phi_{j\alpha}})\bigg)\\
   \D=\sum_{a,b,j,\alpha}
    \frac 1{\lambda-\lambda_\alpha}
    (L^{k-1}(\lambda))_{aj}\bar\phi_{a\alpha}
    \frac 1{\mu-\lambda_\alpha}(L^{l-1}(\mu))_{jb}\phi_{b\alpha}\\
   \D-\sum_{a,b,j,\alpha}\frac 1{\mu-\lambda_\alpha}(L^{l-1}(\mu))_{aj}\bar\phi_{a\alpha}
   \frac 1{\lambda-\lambda_\alpha}(L^{k-1}(\lambda))_{jb}\phi_{b\alpha}\\
   \D=\sum_{a,b,\alpha}\frac 1{\lambda-\lambda_\alpha}\frac 1{\mu-\lambda_\alpha}\bar\phi_{a\alpha}\phi_{b\alpha}
    [L^{k-1}(\lambda),L^{l-1}(\mu)]_{ab}\\
   \D=\sum_{a,b,\alpha}\frac 1{\mu-\lambda}\bigg(\frac 1{\lambda-\lambda_\alpha}-\frac 1{\mu-\lambda_\alpha}\bigg)
    \bar\phi_{a\alpha}\phi_{b\alpha}\cdot\\
   \D\qquad\cdot[L^{k-1}(\lambda),L^{l-1}(\mu)]_{ab}\\
   \D=\frac 1{\mu-\lambda}\tr\bigg((L(\lambda)-L(\mu))
    [L^{k-1}(\lambda),L^{l-1}(\mu)]\bigg)=0.
   \endaligned
\end{equation}
The lemma is proved.
\end{demo}

Using this lemma, we can construct involutive
integrals of motion from $\tr L^k$ $(k=1,2,\cdots)$.

For any complex number $\xi$, let
\begin{equation}
   \det(\xi-L(\lambda))=\xi^3-p_1(\lambda)\xi^2
   +p_2(\lambda)\xi-p_3(\lambda)
\end{equation}
then $p_k(\lambda)$ is the sum of all the determinants of the
principal submatrices of $L(\lambda)$ of order $k$. Suppose the eigenvalues of
$L(\lambda)$ are $\nu_1(\lambda),\nu_2(\lambda),\nu_3(\lambda)$, then
\begin{equation}
   \aligned{l}
   \D\tr L^k(\lambda)=\sum_{j=1}^3 \nu_j^k(\lambda)\\
   \D p_k(\lambda)=\sum_{1\le j_1<\cdots<j_k\le 3}\nu_{j_1}(\lambda)
    \cdots\nu_{j_k}(\lambda).
   \endaligned
\end{equation}
Since $p_k(\lambda)$ can be expressed as a polynomial of $\tr
L^l(\lambda)$ $(l=1,2,\cdots)$, $\{p_j(\lambda), p_k(\mu)\}=0$ for
any two complex numbers $\lambda$ and $\mu$ and two integers
$j,k\ge 0$.

Expand $p_k(\lambda)$ as a Laurant series of $\lambda$
\begin{equation}
   p_k(\lambda)=\sum_{m=-1}^\infty E_m^{(k)}\lambda^{-m-1}
\end{equation}
which is convergent when
$\D|\lambda|>\max_{1\le\alpha\le N}|\lambda_\alpha|$, then
\begin{equation}
   \aligned{l}
   E_m^{(1)}=\innerp 11m
    +\innerp 22m
    +\innerp 33m\\
   E_m^{(2)}=-\innerp 11m
    +\innerp 22m\\
   \qquad+\D\sum_{1\le i<j\le 3}\sum_{l=1}^m\left|\begin{array}{cc}
    \innerp ii{l-1} &\innerp ji{m-l}\\
    \innerp ij{l-1} &\innerp jj{m-l}
    \end{array}\right|\\
   E_m^{(3)}=-\innerp 33m\\
   \qquad-\D\sum_{l=1}^{m}\left|\begin{array}{cc}
     \innerp 11{l-1}
     &\innerp 31{m-l}\\
     \innerp 13{l-1}
     &\innerp 33{m-l}\end{array}\right|\\
    \qquad+\D\sum_{l=1}^{m}\left|\begin{array}{cc}
     \innerp 22{l-1}
     &\innerp 32{m-l}\\
     \innerp 23{l-1}
     &\innerp 33{m-l}\end{array}\right|\\
   \qquad\D+\!\!\!\!\!\!\!\!
   \sum_{\scriptstyle i+j+k=m-2\atop\scriptstyle i,j,k\ge 0}
   \!\!\left|\!\!\begin{array}{c}
   \innerp 11i\\
   \innerp 12i\\
   \innerp 13i
   \end{array}\!\!\!\!\!\!
   \begin{array}{c}
   \innerp 21j\\
   \innerp 22j\\
   \innerp 23j
   \end{array}\!\!\!\!\!\!
   \begin{array}{c}
   \innerp 31k\\
   \innerp 32k\\
   \innerp 33k
   \end{array}\!\!\right|.
   \endaligned\label{eq:E_DS}
\end{equation}
The sums are zero if the lower bound is greater than the
upper bound. These $E_m^{(k)}$'s are in involution.

Let
\begin{equation}
   \aligned{l}
   \Omega_1=\innerp 110=
   (E_0^{(1)}-E_0^{(2)}+E_0^{(3)})/2\\
   \Omega_2=\innerp 220=
   (E_0^{(1)}+E_0^{(2)}+E_0^{(3)})/2\\
   \Omega_3=\innerp 330=-E_0^{(3)}
   \endaligned\label{eq:Omega}
\end{equation}
then
\begin{equation}
   \aligned{l}
   H^x=-E_1^{(1)}-E_1^{(3)}-(\Omega_1-\Omega_2)\Omega_3\\
   H^y=E_1^{(2)}-\Omega_1\Omega_2-\Omega_1\Omega_3-\Omega_2\Omega_3\\
   H^t=-2E_2^{(2)}+2(\Omega_1+\Omega_2)E_1^{(1)}\\
   \qquad +(\Omega_1+\Omega_2-2\Omega_3)H^x+(\Omega_1-\Omega_2)H^y
   \endaligned
\end{equation}
which have all been expressed as polynomials of $E_m^{(k)}$'s.

Hence the following lemma holds.

\begin{lemma}
\begin{equation}
   \{H^x,H^y\}=\{H^x,H^t\}=\{H^y,H^t\}=0,
\end{equation}
\begin{equation}
   \{H^x,E_m^{(k)}\}=\{H^y,E_m^{(k)}\}=\{H^t,E_m^{(k)}\}=0
\end{equation}
and
\begin{equation}
   \{E_m^{(k)},E_p^{(l)}\}=0
\end{equation}
for all $k,l=1,2,3$; $m,p=0,1,2,\cdots$.
\end{lemma}

Now we consider the independence of $E_m^{(k)}$'s.

\begin{lemma}
$E_m^{(k)}$ $(1\le k\le 3;\,0\le m\le N-1)$ are
functionally independent in a dense open subset of
$\hr^{6N}$.
\end{lemma}

\begin{demo}
Let $P_0$ in $\hr^{6N}$ be given by $\phi_{j\alpha}=\epsilon$
$(j=1,2,3; \,\alpha=1,2,\cdots,N)$ where $\epsilon$ is a
small real constant. Then, at $P_0$,
\begin{equation}
   \frac{\partial E_m^{(k)}}{\partial\bar\phi_{j\alpha}}
   =\varepsilon\lambda_\alpha^mb_{kj}+O(\varepsilon^3)
\end{equation}
where
\begin{equation}
   (b_{kj})=\pmatrix{ccc} 1 &1 &1\\ -1 &1 &0\\ 0 &0 &-1\endpmatrix.
\end{equation}
The Jacobian determinant
\begin{equation}
   \aligned{l}
   \D J\equiv\frac{\partial(E_0^{(1)},\cdots,E_{N-1}^{(1)},
    E_0^{(2)},\cdots,E_{N-1}^{(2)},
    E_0^{(3)},\cdots,E_{N-1}^{(3)})}
   {\partial(\bar \phi_{11},\cdots,\bar\phi_{1N},
    \bar\phi_{21},\cdots,\bar\phi_{2N},
    \bar\phi_{31},\cdots,\bar\phi_{3N})}\\
   \D\quad=-2\epsilon^{3N}\left(
   \prod_{1\le \alpha<\beta\le N}
   (\lambda_\beta-\lambda_\alpha)\right)^3
   +O(\epsilon^{3N+2}).
   \endaligned
\end{equation}
$J\ne 0$ at $P_0$ if $\epsilon\ne 0$ is small enough. Since
$J$ is a real analytic function of
$$(\phi_{11},\cdots,\phi_{3N},
\bar\phi_{11},\cdots,\bar\phi_{3N}),$$
$J\ne 0$ in a dense open subset of $\hr^{6N}$. Therefore,
the Jacobian determinant of
$$(E_0^{(1)},\cdots,E_{N-1}^{(1)},
    E_0^{(2)},\cdots,E_{N-1}^{(2)},
    E_0^{(3)},\cdots,E_{N-1}^{(3)})$$
to $6N$ real coordinates
$$(\re(\phi_{11}),\cdots,\re(\phi_{3N}),
\im(\phi_{11}),\cdots,\im(\phi_{3N}))$$
is of full rank $3N$. The lemma is proved.
\end{demo}

In summary, from Lemma 2--5, we have

\begin{theorem}\label{thm:int_Nwave}
The Hamiltonian systems given by the Ha\-mil\-to\-nians (\ref{eq:HBarg})
are involutive and completely integrable in Liouville sense. The
integrals of motion are $E_m^{(k)}$ $(k=1,2,3;\, m=0,\cdots,N-1)$
given by (\ref{eq:E_DS}), which are in involution and functionally
independent in a dense open subset of the phase space $\hr^{6N}$.
Moreover, each solution of these Hamiltonian systems gives a
solution of the DSI equation.
\end{theorem}

\section*{Acknowledgements}
This work was supported by the Chinese Major State Basic Research
Project ``Nonlinear Science'', the City University
of Hong Kong (Grant No.\ 7001041), the Research Grants Council of
Hong Kong (Grant No.\ 9040395,9040466), the Doctoral Program
Foundation and the Key Project for Young Teachers of the Ministry
of Education of China. The first author (Z.~X.~Zhou) is grateful
to the Department of Mathematics of the City University of Hong
Kong for the hospitality.

\end{document}